\documentclass[9pt]{article}
\usepackage{spconf}
\usepackage{dsfont, amsmath,amsthm,amssymb,url,mathtools,algorithmic,algorithm, bbold}
\usepackage{subcaption, mwe,epsfig,graphicx,color}
\usepackage[font=small,labelfont=bf]{caption}

\theoremstyle{definition}

\newcommand{\hanwei}[1]{{\color{black}{{#1}}}}

\DeclareMathOperator*{\argmin}{arg \ min}

\begin{document}
\graphicspath{{figures/}}
\title
{Learning Product Codebooks using Vector-Quantized Autoencoders for Image Retrieval
}

\name {Hanwei Wu and Markus Flierl} 
\address{School of Electrical Engineering and Computer Science\\
KTH Royal Institute of Technology, Stockholm\\
	\{hanwei, mflierl\}@kth.se}
\maketitle
\vspace{1em}
\begin{abstract}
Vector-Quantized Variational Autoencoders (VQ-VAE)\cite{oord:17:nips} provide an unsupervised model for learning discrete representations by combining vector quantization and autoencoders. \hanwei{In this paper, we study the use of VQ-VAE for representation learning for downstream tasks, such as image retrieval. We first describe the VQ-VAE in the context of an information-theoretic framework. We show that the regularization term on the learned representation is determined by the size of the embedded codebook before the training and it affects the generalization ability of the model. As a result, we introduce a hyperparameter to balance the strength of the vector quantizer and the reconstruction error}. By tuning the hyperparameter, the embedded bottleneck quantizer is used as a regularizer that forces the output of the encoder to share a constrained coding space such that learned latent features preserve the similarity relations of the data space. \hanwei{In addition, we provide a search range for finding the best hyperparameter.} Finally, we incorporate the product quantization into the bottleneck stage of VQ-VAE and propose an end-to-end unsupervised learning model for the image retrieval task. The product quantizer has the advantage of generating large-size codebooks. Fast retrieval can be achieved by using the lookup tables that store the distance between any pair of sub-codewords. \hanwei{State-of-the-art retrieval results are achieved by the learned codebooks}.
\end{abstract}
\section{Introduction}	\label{sec:Intro}
Recent advances in variational autoencoders (VAE) provide new unsupervised approaches to learn the hidden structure of data \cite{kingma:14:iclr}. The VAE is a powerful generative model which allows inference of learned latent representations. However, the classic VAEs are prone to the phenomenon of \textquotedblleft posterior collapse \textquotedblright. Here, the latent representations are largely ignored by a decoder that is \textquotedblleft  too powerful \textquotedblright. Vector-quantized variational autoencoders (VQ-VAE) learn discrete representations by incorporating the idea of vector quantization (VQ) into the bottleneck stage. With that, the \textquotedblleft posterior collapse \textquotedblright can be avoided \cite{oord:17:nips}, and the latent features learned by the VQ-VAE are more meaningful.

\hanwei{In the following, we study the use of VQ-VAE for representation learning for downstream tasks, such as image retrieval.} We first describe the VQ-VAE from an information-theoretic perspective by using the so-called variational information bottleneck principle \cite{alemi:17:iclr} \cite{Strouse17}. \hanwei{We show that the regularization term of the latent representation is determined by the size of the embedded codebook, and hence, will affect the generalization ability of the model. Since the regularizer is absent during the training, we introduce a hyperparameter to balance the strength of the vector quantizer and the reconstruction error.} In this way, the bottleneck vector quantizer is used as a regularizer that enforces a constrained code space onto the output of the encoder. This is critical for applications such as image retrieval which require learned latent features that preserve the similarity relation of the input data.  

We further modify the VQ-VAE by introducing a product quantizer (PQ) into the bottleneck stage such that the product codebook can be learned in an end-to-end fashion. Compared to classic vector quantization, the product quantizer can generate an exponentially large codebook at very low memory cost \cite{Jegou:11}. In addition, distance calculations between query and database items in the retrieval process can be avoided by using lookup tables which store the distances between codewords. 
\section{Related Work} \label{sec:back}
Several works have studied the end-to-end discrete representation learning model with different incorporated structures in the bottleneck stages. \cite{theis:17:iclr} and \cite{Balle:17:iclr}  introduce scalar quantization in the latent space and optimize jointly the entire model for rate-distortion performance over a database of training images. \cite{agustsson:17:nips} proposes a compression model by performing vector quantization on the network activations. The model uses a continuous relaxation of vector quantization which is annealed over time to obtain a hard clustering. \cite{jang:17:iclr} and \cite{maddison:17:iclr} introduce the Gumbel-Softmax gradient estimator for non-differentiable discrete distributions. The Gumbel-Softmax estimator determines the gradient of discrete distributions by sampling from a differentiable Gumbel-Softmax distribution which can be smoothly annealed into a categorical distribution. 

For extended works on VQ-VAE, \cite{roy:18} uses the Expectation Maximization algorithm in the bottleneck stage to train the VQ-VAE and to achieve improved image generation results. We note that the authors in \cite{kaiser:18} also explore the product quantization idea for the VQ-VAE and use it to parallelize the decoding process for the sequence model. This approach is known as the decomposed VQ-VAE.  
 \hanwei{\section{Variational Information Bottleneck \\
 		and VQ-VAE}}
 \label{sec: theory}
 \hanwei{Let $I$ denotes the index of the input data, $\mathbf{X}$ the feature representation of the input data, and $Z$ the index of the latent codeword. The objective is to learn a distribution $p(Z|I)$ from the given data distribution $p(I, \mathbf{X})$. Under some constraints, the learned representation $Z$ should retain as much information about $\mathbf{X}$ from $I$ as possible.} The above variables are subject to the Markov chain constraint
 \begin{equation}
 	\mathbf{X} \leftrightarrow I \leftrightarrow Z.
 \end{equation}
   
   \hanwei{Similar to the deterministic information bottleneck (DIB) principle as introduced in \cite{Strouse17}, we focus on minimizing the \emph{representational cost} $H(Z)$ of the learned latent representation. We can formulate the problem  as a rate-distortion-like problem }
 \begin{equation}
 \min_{p(Z|I): d_{\text{IB}}(I, Z)\leq D} H(Z), 
 \end{equation}
 \hanwei{where $d_{\text{IB}}(\cdot)$ is the information bottleneck distortion and $D$ is a constant constraint on the information bottleneck distortion}. Similar to rate-distortion optimization, the objective function can be expressed by the equivalent Lagrangian formulation
 \begin{equation}
 L_{\text{IB}} = d_{\text{IB}}(I, Z) + \mu H(Z),
 \end{equation}
 where $\mu$ is the Lagrangian parameter.
 
Now, consider the case where the information bottleneck distortion is defined as the \hanwei{Kullback-Leibler (KL) divergence between the true data distribution and the data distribution generated by the latent representation\cite{tishby:03}, \it{i.e.},}
 \begin{equation}
 \label{eq: IB_obj}
 	d_{\text{IB}}(I, Z) = \text{KL}(p(\mathbf{X}|I)\|p(\mathbf{X}|Z)).
 \end{equation}
We can decompose $d_{\text{IB}}(I, Z)$ into two terms 
\begin{align}
 &\text{KL}(p(\mathbf{X}|I)\|p(\mathbf{X}|Z)) \\
 = &\sum_{i}\sum_{z} p(i)p(z|i) \int p(\mathbf{x}|i)\log \frac{p(\mathbf{x}|i)}{p(\mathbf{x}|z)}d\mathbf{x} \\
\label{eq: joint}
= &\int\sum_{z}p(\mathbf{x}, z)\log \frac{p(z)}{p(\mathbf{x}, z)}d\mathbf{x} -
\int\sum_ip(i, \mathbf{x})\log \frac{p(i)}{ p(i, \mathbf{x})}d\mathbf{x},
\end{align}
where (\ref{eq: joint}) is derived from using the chain rule to express the conditional probability $p(\mathbf{x}|z)$ as
\begin{equation}
p(\mathbf{x}|z) = \frac{1}{p(z)}\sum_i p(\mathbf{x}|i)p(z|i)p(i).
\end{equation}

The second term of (\ref{eq: joint}) is determined solely by the given data distribution $p(I, \mathbf{X})$ and is a constant. Hence, it can be ignored in the loss function for the propose of minimization. The first term of (\ref{eq: joint}) can have an upper bound by replacing the $p(\mathbf{x}|z)$ with a variational approximation $q(\mathbf{x}|z)$ \cite{alemi:17:iclr}
\begin{align}
&\int\sum_{z}p(\mathbf{x}, z)\log \frac{p(z)}{p(\mathbf{x}, z)}d\mathbf{x} \\
=&-\sum_z p(z)\int p(\mathbf{x}|z)\log p(\mathbf{x}|z) d\mathbf{x}\\
\label{ieq: 1}
 \leq& -\sum_z p(z)\int p(\mathbf{x}|z)\log q(\mathbf{x}|z) d\mathbf{x} \\
\label{eq:reconstruct}
 =& -\sum_i p(i) \int p(\mathbf{x}|i)\sum_{z}p(z|i)\log q(\mathbf{x}|z) d\mathbf{x},
\end{align}
 where (\ref{ieq: 1}) results from the nonnegativity of the KL divergence
 \begin{align}
 	\text{KL}(p(\mathbf{X}|Z)\|q(\mathbf{X}|Z)) &\geq 0    \\
 	\int p(\mathbf{x}|z)\log p(\mathbf{x}|z) d\mathbf{x} &\geq \int p(\mathbf{x}|z) \log q(\mathbf{x}|z)d\mathbf{x}.
 \end{align}
 
 \hanwei{In the VQ-VAE setting, the vector quantization is performed on the output of the encoder. Hence, the input of the decoder is the closest codeword in the codebook. The size of the codebook is prespecified. The conditional probability distribution $p(Z|I)$ can be considered as being parameterized by the a encoder neural network $f(\cdot)$ with a nearest neighbor search on the codebook  
 	\[
 	p(z = k|I = i) = \left\{
 	\begin{array}{ll}
 	1  \hspace{1em}\text{for} \ k = \underset{z \in [K]}{\argmin} \|z_e(\mathbf{x})-e_z\|_2,\\
 	0 \hspace{1em}\text{otherwise}
 	\end{array},
 	\right.\]
 	where $z_e$ is the output of the encoder network $f(\cdot)$, $K$ is the number of codewords of the quantizer, and $e_j, j = 1, \dots, K$ is the codeword. The conditional distribution $q(\mathbf{X}|Z)$ can be considered as being parameterized by the decoder neural network $g(\cdot)$ and a codeword lookup function $Q(\cdot)$ that maps the index to the codeword
 \begin{equation}
 q(\mathbf{x}|z) = g(\mathbf{x}|Q(z)) =  g(\mathbf{x}|z_q),
 \end{equation}
  where we use $z_q = e_z$ to represent the input of the decoder.
 
 Since $p(\mathbf{z}|i)$ is a one-hot vector and the decoder should not allocate any probability mass to $g(\mathbf{x}|e_z)$ for $e_z \neq z_q(\mathbf{x})$, we can write that 
 \begin{equation}
 \sum_{z}p(z|I = i)\log q(\mathbf{x}|z) = \log g(\mathbf{x}|z_q(\mathbf{x}_i)).
 \end{equation}
  If we assume that $g(\mathbf{x}|z_q(\mathbf{x}_i)) = \mathcal{N}(\mathbf{x}|\hat{\mathbf{x}}, \mathbf{1})$, with $\hat{\mathbf{x}} = z_q(\mathbf{x}_i)$, then the log likelihood of $q(\mathbf{x}|z_q(\mathbf{x}_i))$ is proportional to the squared difference between the input and the output of the decoder. Therefore, (\ref{eq:reconstruct}) becomes the mean square error between the input and output which is considered as the reconstitution error of the model.}
 
\hanwei{For the regularization term in (\ref{eq: IB_obj}),} we can use a similar variational approximation technique to upper bound the entropy $H(Z)$ of the latent variable as \cite{Strouse17}
 	  \begin{align}
 	  \label{ieq:2}
 	  H(Z) &\leq -\sum_{z}p(z) \log r(z)  \\
 	  &= -\sum_{i}\sum_{z}p(i)p(z|i)\log r(z)  \\
 	   \label{eq:crossentropy}
 	  & = -H(p(Z), r(Z))
 	  \end{align}
where (\ref{ieq:2}) results from the nonnegativity of the KL divergence.
\begin{equation}
	\text{KL}(p(Z) \| r(Z)) \geq 0 \ \rightarrow \ \sum_z p(z)\log p(z) \geq \sum_z p(z)\log r(z).
\end{equation}
As a standard practice, the marginal $r(Z)$ is set to be a simple uniform distribution. Then the cross entropy  (\ref{eq:crossentropy}) becomes a constant equal to $\log K$ and can be omitted from the loss function. \hanwei{That is, the constraint on the learned representations is determined by the size of the embedded codebook before the training. Therefore,  a tractable upper bound for (\ref{eq: IB_obj}) becomes the reconstruction error between the input and output of the model.

Since quantization is not a continuous function, it has no gradient for the backpropagation training. \hanwei{The VQ-VAE solves the problem by using stop gradient operations to 
 	   	  		optimize the encoder-decoder and the bottleneck vector quantizer separately.} The stop gradient operator can output its input when it is in the forward pass and does not take it into account when computing gradients in the training process. The strategy is to copy the gradients that optimize the reconstruction loss at the input of the decoder to the output of the encoder. This can be easily achieved by representing the input of the decoder as 
    	  		\begin{equation}
    	  		z_q = z_e + \text{sg}(z_q-z_e),
    	  		\end{equation}
    	  		 where $\text{sg}(\cdot)$ denotes the stop gradient operator. As a result, codewords receive no update gradients. Instead, the VQ-VAE adds a second component, $\|\text{sg}(z_e)-z_q\|_2^2$, to optimize the codewords. In this way, codewords can move to the output of encoder which is fixed during the optimization. 
    	  		 
    	  		 Therefore, we can recover the empirical loss function of the VQ-VAE \cite{oord:17:nips} from (\ref{eq:reconstruct}) as}
    	  		 \begin{equation}
    	  		 \label{eq:loss1}
    	  		 \begin{split}
    	  		 &L_{\text{VQ-VAE}} = \frac{1}{N}\sum_{i = 1}^{N}\left[\log g(\mathbf{x}|z_q(\mathbf{x}_i))+ \|\text{sg}\left(z_e(\mathbf{x}_i)\right)-z_q(\mathbf{x}_i)\|_2^2 \right.\\
    	  		 &\left.  + \beta\|z_e(\mathbf{x}_i)- \text{sg}\left(z_q(\mathbf{x}_i)\right)\|_2^2 \right],
    	  		 \end{split}
    	  		 \end{equation}
    	  		 \hanwei{where we assume $p(i) = \frac{1}{N}$. A third term is added to the objective function of (\ref{eq:loss1}) as the commitment loss. It is used to force the encoder output $z_e(\mathbf{x})$ to commit to a codeword. $\beta$ is a constant weight parameter for the commitment loss.}
 	  \vspace{1em}
 	 \section{Bottleneck Vector Quantizer as a Regularizer}
 	 \label{sec:vqregularizer}
 	  \subsection{Effects of the Bottleneck Vector Quantizer}
\begin{figure}[!h]
    	\centering
    	\includegraphics[width=\linewidth]{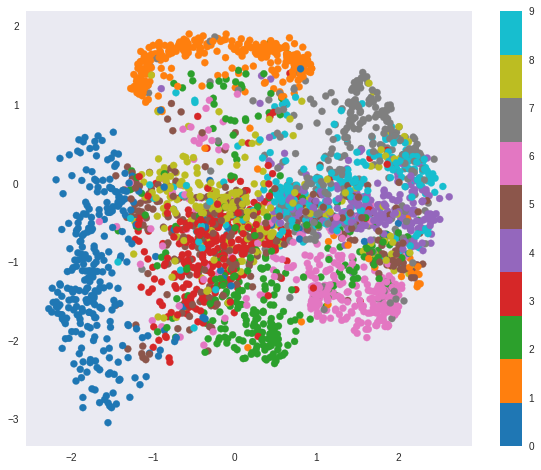}
    	\captionof{figure}{Number of codewords $K = 20$.}
    	\label{fig:k=20}  	 	
    \end{figure}

\begin{figure}[!h]
    		\centering
    		\includegraphics[width=\linewidth]{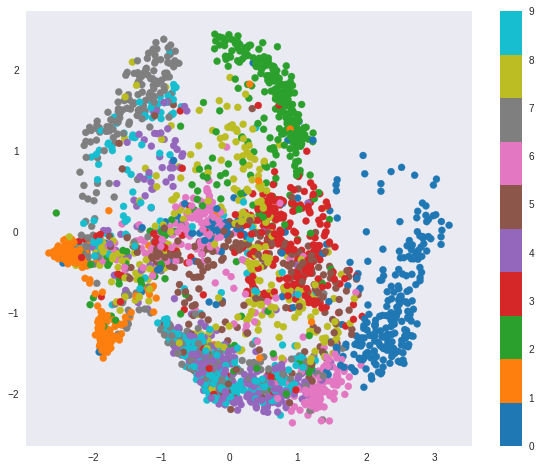}
    		\captionof{figure}{Number of codewords $K = 25$.}
    		\label{fig:k=20a}  	 	
    	\end{figure}

 	   \hanwei{According to \cite{zhao2017infovae} and \cite{burgess2018understanding}, the generalization ability of the model can be indicated by the meaningfulness of the learned representations. The meaningful representation should preserve the similarity relations of the data space. The high rate setting of the vector quantizer increases the learning capacity of the model and tends to overfit the data. Since the output of the encoder has more codeword choices, for easier decoding, it is more likely to be quantized into codewords that are far away from each other. Although a lower reconstruction error can be achieved due to the high discriminability of latent codewords, the generalization ability is poor in this case. On the other hand, the low rate setting decreases the average discriminability of the input data. As a result, the reconstruction increases as the rate decreases. However, in order to achieve lower reconstruction errors for the low rate setting, the model is forced to ensure that neighboring data points are also represented closely together in the latent space, which leads to a better generalization ability. 
 	   	
 	   	In Figures \ref{fig:k=20}  and \ref{fig:k=20a}, we plot the two-dimensional latent representations learned from the MNIST dataset using different sizes of embedded codebooks. Different colors indicate different digit classes. We use an encoder network with layer structure setting as $L$-$500$-$500$-$2000$-$2$ and an decoder network with layer structure setting as $2$-$2000$-$500$-$500$-$L$ to learn $2$-dimensional latent representations, where $L$ is the input data dimension. All layers are fully connected and use rectified linear units ReLUs as activation functions. Although the high rate case $(K = 25)$ has lower reconstruction error than the low rate case $(K = 15)$, we can observe from the figures that the latent representation of different digit classes for the high rate case are more overlapped with each other than for the low rate case. This indicates the bad generalization ability of the high rate setting.}
    	
 	   \hanwei{
 	   	In order to solve the problem of the bad generalization ability for the high rate setting, we introduce a hyperparameter to control the strength of the vector quantizer. Specifically, we introduce a multiplicative weight $\lambda$ to both second and third terms of (\ref{eq:loss1}) to control the updating power of the vector quantizer.} If we increase the value of $\lambda$ for a fixed rate, the vector quantizer becomes ``more powerful". This minimizes the quantization error and pushes codewords far away from each other. This may decrease the reconstruction error, but leads to a bad generalization ability. In this case, the input data is less likely to be updated to another codeword due to the weak encoder-decoder optimization in the first term of (\ref{eq:loss1}). On the other hand, a small value of $\lambda$ creates a ``weaker" vector quantizer \hanwei{and the quantization error increases. This is equivalent to adding noise to the latent codewords. In this case, the input data will be easily swayed away to other codewords due to the increased quantization error. This creates similar effects as the low rate setting of the vector quantizer, where the locality of the data space in the latent space is better preserved. In this way, the bottleneck vector quantizer is used as a regularizer of the latent representation as it enforces a shared coding space on the encoder output.
    	
    	In Figure \ref{fig:k=20b}, we plot the learned representation using the same layer structures as in Figures \ref{fig:k=20} and \ref{fig:k=20a}, but using the introduced hyperparameter $\lambda = 0.5$. We can observe that the latent representation preserves the similarity relations of the input data better than the representation as shown in Figure \ref{fig:k=20a} in terms of using the space more uniformly..
    	\begin{figure}[!h]
    	\centering
    	\includegraphics[width=\linewidth]{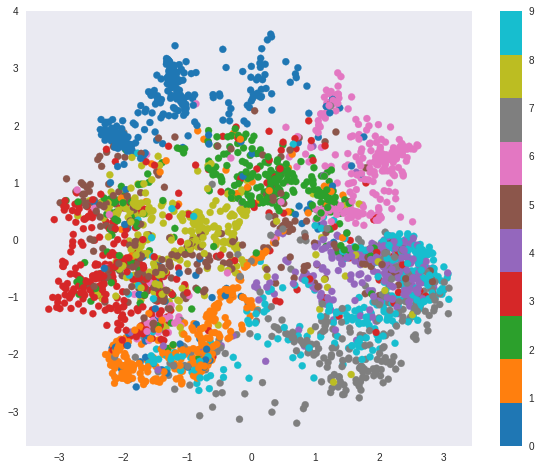}
    	\captionof{figure}{Number of codewords $K = 25$, hyperparameter $\lambda$ = 0.5.}
    		\label{fig:k=20b}  	 	
    \end{figure}}
 	 \subsection{The Tuning of the Hyperparameter $\lambda$}
 	  	 \hanwei{Since the appropriate values of $\lambda$ depend on the model,} we use the ratio of the average distances of the closest codeword to the second closest codeword of the input data as a robust indicator for the ``impact" of the vector quantizer. In Figure \ref{fig:test2}, we plot the reconstruction error over above distance ratio for varying the values of $\lambda$. We observe that the reconstruction error decreases for distance ratios of $0.35$ to $0.6$ as the codewords are updated more easily by the encoder-decoder optimization. However, when the distance ratio exceeds approximately $0.62$, the input data is swayed over too easily to other codewords by the encoder-decoder optimization. \hanwei{Therefore, the optimization becomes more difficult to converge} and the reconstruction error increases significantly. \hanwei{In order to find an appropriate value for the hyperparameter $\lambda$, we limit our search of $\lambda$ by restricting the permissible range of distance ratios. 
 	  	 	\begin{figure}[!h]
 	  	 		\centering
 	  	 		\includegraphics[width=\linewidth]{{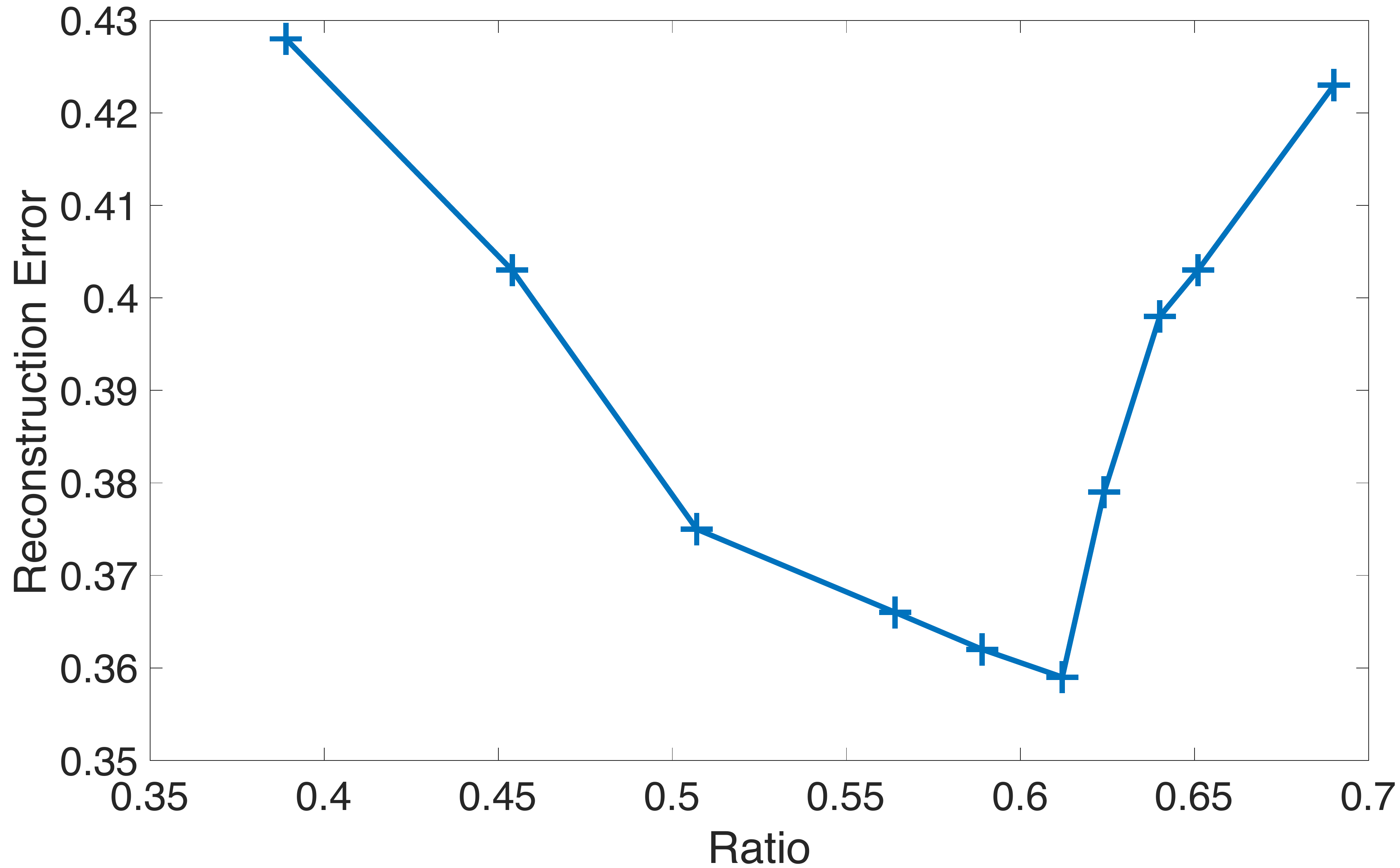}}
 	  	 		\captionof{figure}{Reconstruction error over distance ratio for varying $\lambda$.}
 	  	 		\label{fig:test2}
   	 	\end{figure}}
 	 \section{Application: Image Retrieval} 
 	 \label{sec: application}
 	 \subsection{PQ-VAE}
 	 \begin{figure}[!h]
 	 	\centering
 	 	\includegraphics[width=0.5\textwidth]{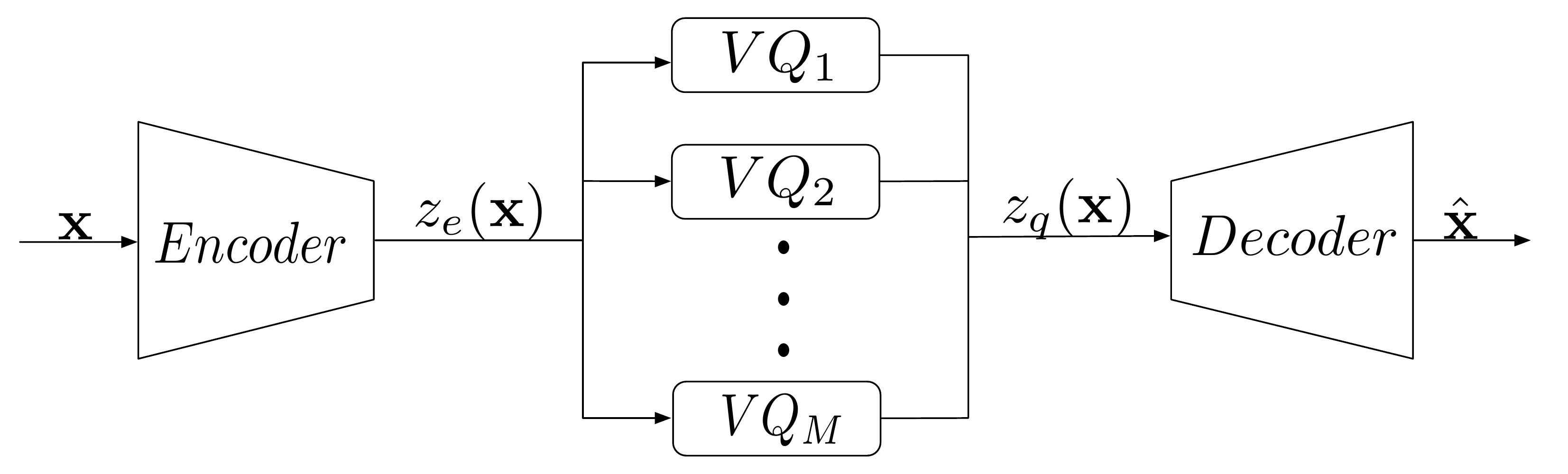}
 	 	\caption{System Model for PQ-VAE.}
 	 	\label{fig:sys1}
 	 \end{figure}
 	 The model of PQ-VAE consists of an encoder, a decoder, and a bottleneck product quantizer. The encoder learns a deterministic mapping and outputs the latent representation $z_e(\mathbf{x}) \in \mathbb{R}^{D}$, where $\mathbf{x} \in \mathbb{R}^{L}$ denotes the input data. The learned latent representation $z_e(\mathbf{x})$ can be seen as an efficient representation of the input data $\mathbf{x}$, such that $D \ll L$. 
 	 
 	 The latent representations $z_e(\mathbf{x})$ are then fed into a product quantizer. The product quantizer consists of $M$ sub-vector quantizers which handle sub-vectors of dimension $D/M$. Each $\text{VQ}_m$ partitions the subspace into $K$ clusters. The sub-clusters are characterized by the sub-codebook $C^{(m)} = \{e_1^{(m)}, \cdots, e_K^{(m)}\}$, where $m = 1, \cdots, M$ denotes the m-th sub-vector quantizer.. 
 	 
 	 Each sub-vector is quantized to one of the $K$ codewords by the nearest neighbor search
 	 \begin{equation}
 	 \label{eq:assgn1}
 	 z_q^{(m)}(\mathbf{x}) = e_k^{(m)}, \ \text{where} \ k = \argmin_i\|z_e^{(m)}\left(\mathbf{x}\right)-e_i^{(m)}\|_2.
 	 \end{equation} 
 	 The output $z_q^{(m)}(\mathbf{x})$ of the M sub-quantizers is concatenated to the full codeword $z_q(\mathbf{x}) = \left[ z_q^{(1)}(\mathbf{x}), z_q^{(2)}(\mathbf{x}), \cdots, z_q^{(M)}(\mathbf{x})\right]$ and then passed as input to the decoder. The decoder then reconstructs the input images $\mathbf{x}$ given the full codewords $z_q(\mathbf{x})$. Figure \ref{fig:sys1} shows the whole system model.
 	  	 
 	  We modified the loss function of VQ-VAE from \cite{oord:17:nips} to adapt to the product quantization setting as
 	 \begin{align}
 	 \label{eq:loss2}
 	 L =& -\log g(\mathbf{x}|z_q(\mathbf{x})) + \lambda\left(\sum_{m = 1}^{M}\beta\|z_e^{(m)}(\mathbf{x})\right.\\ \nonumber
 	 &\left.- sg\left(e^{(m)}\right)\|_2^2 + \sum_{m = 1}^{M}\|\text{sg}\left(z_e^{(m)}(\mathbf{x})\right)-e^{(m)}\|_2^2\right), 
 	 \end{align}
 	 where $\lambda$ is the introduced regularization weight for the used vector quantizer. 
 	 
 	 The $M$ sub-quantizers are trained simultaneously and independently. The codeword is simply updated by the average of the latent values that have been assigned to each cluster in each iteration. For the $i$-th codeword of the $m$-th sub-quantizer, the $i$-th sub-codeword is updated by the following rules:
 	 \begin{align}
 	 	n_i &:= \sum_{j = 1}^{B} \mathbb{1}(z_j = i)\\
 	 	e_i &:= \frac{1}{n_i}\sum_{j = 1}^B\mathbb{1}(z_j = i)z(x_j),
 	 \end{align}
 	 	 where $\mathbb{1}(\cdot)$ is the indicator function, and $B$ is the size of the mini-batch. The codeword assignment of the latent representation follows the nearest neighbor search as in (\ref{eq:assgn1}).
 	 \subsection{Querying}
 	  The discrete encoding $\mathbf{z}$ of the input images can be generated using the trained encoder and learned product codebook. For each input image, $N$ discrete encodings are generated by the encoder. Hence, each image is represented by an encoding vector $\mathbf{z} = \left[z_1^{(1)}, \cdots, z_1^{(M)}, \cdots, z_N^{(M)}\right]$ of size $M \times N$, where each element is the index of its quantized sub-codeword
 	 \begin{equation}
 	 \label{eq:nn1}
 	 z^{(m)}= \arg \min_{i \in [K]} \|z_e^{\left(m\right)}(\mathbf{x}) -e_i^{(m)}\|_2.
 	 \end{equation}
 	 In this way, an input image can be compressed to a code of length
 	 \begin{equation}
 	 \label{eq:bitscalculate}
 	 R = NM\log_2 K.
 	 \end{equation}
 	 
\begin{figure}[!h]
 	 	\centering
 	 	\includegraphics[width=0.5\textwidth]{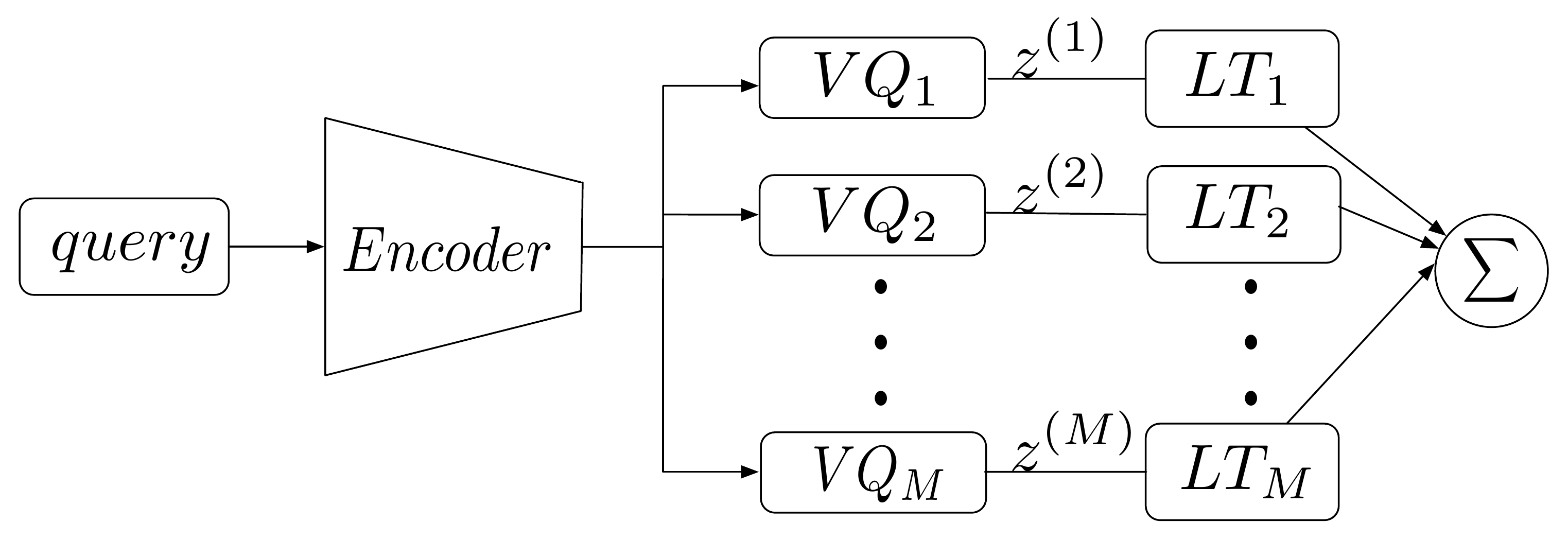}
 	 	\caption{Querying process.}
 	 	\label{fig:rd}
 	 \end{figure}

 	 The querying is conducted in the quantized space. Both database and query images are fed into the trained encoder and learned product codebook. We use the discrete encoding of query and database images by the product quantizer for querying and storing. In the database, we store $M$ Lookup Tables (LT) with $K \times K$ entries. Each LT stores the distances between every two sub-codewords of its sub-codebook. The image encodings are used as the indices of the table. When querying, the distance between query $\mathbf{q}$ and database $\mathbf{x}$ is obtained by summing up the distances as given by the LTs
 	 \begin{equation}
 	 	\label{eq: dist1}
 	 	d(\mathbf{q}, \mathbf{x}) = LT_1\left(z_q^{(1)}, z^{(1)}\right) + \cdots + LT_M\left(z_q^{(M)}, z^{(M)}\right),
 	 \end{equation}
 	 where $z_q$ is the encoding of the query image, and $z$ is the encoding of the database image. Hence, fast retrieval can be achieved because no additional distance calculations are needed. Figure \ref{fig:rd} shows the querying process.
 	 \subsection{Results}
 	 We use the CIFAR-10 dataset which contains $60000$ images of size $32\times32\times3$ to test the performance of PQ-VAE on the image retrieval task. We train the model by using $50000$ images from the training set. Further, we treat their discrete encodings as the database items. $10000$ test images are used as queries. We use the mean Average Precision (mAP)  of the top $1000$ returned images as the performance measure. 
 	 
 	  The encoder consists of $3$ strided convolutional layers with stride $2$ and filter size $4 \times 4$, followed by one max pooling layer and two residual $3 \times 3$ blocks (implemented as ReLU, $3\times 3$ conv, ReLU, $1 \times 1$ conv), all having $256$ hidden channels. In this setting, the input images are compressed into $N = 2 \times 2$ discrete encodings. 
 	 
 	 The decoder follows a structure that is symmetric to the encoder. It consists of two residual $3 \times 3$ blocks at the beginning, followed by a resize layer up to the slice size of $4\times4$. Then it is followed by two transposed convolutional layers with stride $2$ and filter size $4 \times 4$, with a resize layer up to the size of $16 \times 16$ in between.
 	 
 	 We use the ADAM optimizer \cite{kingma:15:iclr} with learning rate 2e-4 and evaluate the performance after $25000$ iterations with batch-size $100$.  The decay parameter of $\gamma$ of EMA is set to be $0.99$ as suggested in \cite{oord:17:nips}.
 	 
 	 We test the product quantizer that consists of $M = 4$ sub-quantizers. We set the number of codewords of the sub-quantizers to $K = 2, 8, 16$. According to (\ref{eq:bitscalculate}), this corresponds to rates of $32$, $48$ and $64$ bits for the latent representations. We compare our model to other state-of-the-art methods in Table \ref{table:1}. Our proposed model outperforms the reference methods in the table. Note that although \cite{lin:16:cvpr} and \cite{Song:AAAI:18} give better results than our proposed model for the CIFAR-10 dataset, they use VGG networks \cite{simonyan:14} which are pretrained by using ImageNet in a supervised fashion. On the other hand, our model is completely unsupervised and it is trained from scratch.
 	 \begin{table}[!h]
 	 	\centering
 	 	\begin{tabular}{| l | l | l | l |}
 	 		\hline
 	 		&32 bits & 48 bits & 64 bits \\ \hline
 	 		LSH \cite{indyk:04}&12.00 &12.00&15.07\\ \hline
 	 		Spectral Hashing \cite{Weiss:08:nips}&13.30&13.00&13.89\\ \hline
 	 		Spherical Hashing \cite{heo:15:pami} &13.30  &13.00  &15.38\\  \hline
 	 		ITQ \cite{lazebnik:13:pami}&16.20&17.50&16.64 \\ \hline
 	 		Deep Hashing \cite{jiwen:15:cvpr}& 16.62& 16.80 & 16.69 \\ \hline
 	 		PQ-VAE & $\mathbf{21.86}$ & $\mathbf{22.79}$ & $\mathbf{23.42}$ \\ \hline
 	 	\end{tabular}
 	 	\caption{Mean Average Precision of the top 1000 returned images for compression rates of 32, 48, and 64 bits.} \label{table:1}
 	 \end{table}
 	 \section{Conclusions}
 	 We extended the work of VQ-VAE by embedding a product quantizer into the bottleneck of an autoencoder for the image retrieval task. We formulate the VQ-VAE problem by using the so-called variational information bottleneck principle and show that the regularization term for the learned representations is determined by the size of the embedded codebook. We introduce a hyperparameter to control the impact of the vector quantizer such that we can further regularize the latent representation. \hanwei{With appropriately tuned hyperparameter, we show that our learned representations improve the mean average precision of the investigated image retrieval task.}
\bibliographystyle{IEEEbib}
\bibliography{fine3}
\end{document}